\documentclass[iop]{emulateapj}
\usepackage{natbib}
\usepackage{amsmath, amssymb}
\usepackage{times}
\usepackage{apjfonts, graphicx, graphics, mathtools}
\usepackage[breaklinks,colorlinks,urlcolor=blue,citecolor=blue]{hyperref}
\usepackage[all]{hypcap}
\usepackage{aas_macros}
\usepackage{subfigure}
\usepackage{multirow}
\usepackage{hhline}
\usepackage{lineno}
\usepackage{setspace}
\usepackage{ulem}
\usepackage{cancel}

\shorttitle{The role of hydrostatic pressure in Molecular Cloud Formation}
\shortauthors{Iu. V. Babyk et al.}

\begin{document}

\title{Atmospheric pressure and  molecular cloud formation in early-type galaxies}
\author{Iu. V. Babyk$^{1,2,3,4\ast}$}
\author{B.~R. McNamara$^{2,5}$}
\author{P.~E.~J. Nulsen$^{3,6}$}
\author{H.~R. Russell$^{7}$}
\author{A.~C. Edge$^{8}$}
\author{Leo Blitz$^{9}$}

\affil{
    $^{1}$ Department of Physics and Astronomy, University of California at Irvine, 4129 Frederick Reines Hall, Irvine, CA 92697, USA \\ 
    $^{2}$Department of Physics and Astronomy, University of Waterloo, 200 University Avenue West, Waterloo, ON, N2L 3G1, Canada \\
    $^{3}$ Center for Astrophysics | Harvard-Smithsonian, 60 Garden Street, Cambridge, MA 02138, USA \\
    $^{4}$Main Astronomical Observatory of the National Academy of Sciences of Ukraine, 27 Zabolotnoho str., 03143, Kyiv, Ukraine \\
    $^{5}$Perimeter Institute for Theoretical Physics, Waterloo, ON, N2L 2Y5, Canada \\
    $^{6}$ICRAR, University of Western Australia, 35 Stirling Hwy, Crawley, WA 6009, Australia \\
    $^{7}$Institute of Astronomy, Madingley Road, Cambridge CB3 0HA, UK \\
    $^{8}$ Department of Physics, University of Durham, South Road, Durham DH1 3LE, United Kingdom \\
    $^{9}$ Department of Astronomy, University of California, Berkeley, CA 94720, USA\\
    \\
}

\begin{abstract}
\hspace{0.5cm} {\bf }{A strong correlation between atmospheric pressure and molecular gas mass is found in central cluster galaxies and early-type galaxies.} This trend and a similar trend with atmospheric gas density would naturally arise if the molecular clouds condensed from hot atmospheres.  Limits on the ratio of molecular to atomic hydrogen in these systems exceed unity. The data are consistent with ambient pressure being a significant factor in the rapid conversion of atomic hydrogen into molecules as found in normal spiral galaxies.

\end{abstract}


\altaffiltext{*}{
    \href{mailto:babikyura@gmail.com}{babikyura@gmail.com}
}

\section{Introduction}\label{sec:intro}

Central cluster and elliptical galaxies contain gas over a broad range of temperatures. The most abundant form resides in a hot atmosphere with temperatures of  $10^7$K-$10^8$ K maintained close to hydrostatic equilibrium.  Cooler ionized gas above 10$^{4}$ K \citep{Bregman:06, Crawford:99, McDonald:10, McDonald:11}, atomic  \citep{ODea:98, Welch:10, Oosterloo:10, Putman:12, Westmeier:18} and molecular gas at colder temperatures are present in smaller amounts. \citep{Edge:01, Edge:02, Salome:03,Salome:04, Salome:06, Dona2011,Pulido:17, Russell19, Oliv19}.  

This gas and its phases may originate from several sources including, stellar mass loss, accretion from other galaxies, accretion from the intergalactic medium, and primordial gas remaining from the early moments of galaxy formation \citep{Bregman07, Dekel09}.  Some of this gas may end up in the hot phase as it is heated by shocks to the virial temperature of the halo.
Recent analyses indicate that a significant fraction of the molecular gas in early-type and central cluster galaxies may be produced by  atmospheric cooling from the hot phase gas \citep{Werner19, Babyk:17prof, Pulido:17}.  Absent continual energy input, the hot atmosphere will radiate away its thermal energy and condense onto the host galaxy, fueling both star formation and feedback from nuclear black holes \citep{Fabian94, Fabian:12, McNamara:07, McNamara:12}. This development is significant.  Observational and theoretical considerations indicate that the growth of massive galaxies is both suppressed and regulated by feedback from supermassive black holes and stellar winds \citep{Croton:06, Sijacki:07, Bower06, Behroozi13}.  Self-regulated growth via black hole feedback requires a steady supply of fuel. Self-regulation would be difficult to maintain if cold fuel arrived randomly, as expected for mergers and inflow from the intergalactic medium.  Therefore understanding the origins of all gas phases in galaxies is central to understanding how galaxies themselves have evolved.

Why cold gas in central cluster galaxies is predominantly molecular hydrogen rather than atomic hydrogen is poorly understood. Early searches for HI at the centers of clusters revealed only upper limits \citep{Haynes79, McNamara:90, Odea94}.  The limits were sufficient to exclude the notion that atmospheres were cooling at their radiative rates and settling into the atomic gas.  These surveys detected atomic hydrogen in absorption toward several objects.  Targeting clusters identified by the Rosat All Sky Survey, \citet{Edge:02} and \citet{Salome:03} discovered molecular clouds in great abundance at the centers of  clusters with the highest radiative cooling rates.  The amounts, despite exceeding $10^{10}~\rm M_\odot$, are too low to account for atmospheric cooling at unimpeded rates.  Nevertheless, the relatively high molecular gas masses indicated that hot atmospheres were indeed likely to be cooling significantly with the cooling gas being converted efficiently from atoms to molecules.  

Molecules form primarily on dust grains, even those with relatively high kinetic temperatures \citep{Lebourlot:12}. Unshielded dust would sputter away rapidly in the harsh cluster environment \citep{Draine:79}. Nevertheless, dust is abundant in the star formation and molecular gas complexes located at the centers of clusters \citep{Vant18,Dona2011, Russell19, Oliv19}.  Dust is also abundant in giant elliptical galaxies \citep{Sadler:85, Goudfrooij:94, Vandokkum:95}.  It is usually associated with cold atomic and molecular gas \citep{Combes:07}.  Thirty to forty percent of early-type galaxies contain detectable levels of molecular gas \citep{Combes:07, Salome:11, Young:11}. Infrared and radio observations indicate that a minority of systems with relatively large amounts of molecular gas form stars at rates of $\sim 0.1~\rm M_\odot yr^{-1}$ or so \citep{Combes:07, Shapiro:10, Ford:13}. Despite the harsh atmospheric environment, conditions are evidently ripe for dust and molecular gas formation.  

The efficiency of conversion from HI to  $\rm H_2$ depends critically on the strength of the radiation field and the pressure of the surrounding interstellar medium \citep{Elmegreen:93}. Apparently, even a slight increase in pressure can convert an entire HI cloud into molecular gas. Conversely, increased levels of ultraviolet radiation can quickly dissociate molecular hydrogen into a primarily atomic phase.
Large regions of the interstellar medium may transform quickly into molecules when experiencing increased pressure in spiral arms \citep{Elmegreen:93}.  Molecules may then return to atoms in the higher photon density regions of star formation.  
Studying spiral galaxies with a variety of morphologies, \citet{Blitz:06} and \citet{Leroy:08} found a strong correlation between the surface density ratio of  $\rm H_2$/HI such that the ratio increases well above unity with increasing midplane pressure. This trend has been studied previously only in spiral galaxies.  Here we examine the relationship between hydrostatic pressure and molecular gas abundance in a relatively large sample of central cluster galaxies and normal ellipticals.

\section{Data }\label{sec:2}
\subsection{X-rays}
Data taken from \citet{Hogan:17a, Pulido:17, Babyk:17prof, Babyk:17ent} were used to extract hydrostatic pressure profiles for the objects studied here. These works include 110 clusters and 58 groups and galaxies sufficiently nearby and with enough exposure time to resolve the thermodynamic properties of the inner regions ($<$ 10 kpc) of their atmospheres. Descriptions of the data reduction procedures, spectral fitting, and the calculation of the pressure profiles can be found in \citet{Hogan:17a, Pulido:17, Babyk:17prof, Babyk:17ent}. Here we summarize briefly.

Data reduction, including reprocessing and bad pixel extraction, was performed using the CIAO v.4.2 software package. To create clean, level-2 event files, the {\sc chandra\_repro} tool was applied. The {\sc acis\_process\_events} tool was used to correct the time-dependent gain, while the {\sc lc\_clean} tool provided by M. Markevitch was used to detect and remove the background flares. We used the {\sc wavdetect} tool to identify and remove point sources. Each observation has been processed with the corresponding blank-sky background file. Multiple observations were reprojected to match the observation with the longest exposure time. The X-ray spectra were extracted using circular annuli and then deprojected using the {\sc DSDEPROJ} routine \citep{Russell:08}. The ancillary response and response matrix files were extracted using {\sc mkwarf}
and {\sc mkacisrmf} tools. The chip gaps and the area lost to point sources were corrected with exposure maps.

The projected and deprojected spectra have been fitted using {\sc xspec} v.12.8 software \citep{Arnaud:96}. The spectral modeling has been performed using an absorbed thermal model. The projected and deprojected temperature and spectral normalization were used to measure atmospheric density, which in turn were used to calculate hydrostatic pressure. 

Additionally, we analyzed the small sample of early-type galaxies which have both detections for atomic neutral and cold molecular gas. The whole sample is given in \citet{Welch:10}. We selected 8 targets and analyzed their Chandra data following \citet{Babyk:17scal}. Due to the poor data, we applied $\beta$-model to fit surface brightness profiles and to get modeled X-ray characteristics of hot gas. For two systems we have analyzed XMM-Newton data. We analyzed those data following \citet{Pointec:05} with the latest calibration files and pipeline background subtraction. We extracted and inspected the light curves in 10-12 keV for MOS and 12-15 keV for PN cameras to screen the periods with a high background. We removed periods with the rate of 0.15 count per second for MOS and 0.22 count per second for PN cameras. We extracted the combined, background-subtracted X-ray image in the 0.5-7.0 keV band similar to our Chandra analysis. We generated a blank-sky background event file for each instrument and observation. The total background was then subtracted from the data and final X-ray images were extracted.

For both Chandra and XMM-Newton data we extracted X-ray profiles, including temperature, density, entropy, mass, and pressure profiles. Due to quite large uncertainties in the central bins of XMM features, we used the pressure values at 10 kpc only. We used measurements of additional samples to plot $HI-P$ and $H_2/HI-P$ relations only.

\subsection{Atomic and molecular Gas}

Molecular gas masses were taken from \citet{Pulido:17}, \citet{ Babyk:17prof}, and \citet{Welch:10}. Molecular gas mass measurements were available for 84 targets, including 39 for the low-mass systems (mostly ellipticals). The CO data were obtained from a variety of instruments.  Pulido's sample included IRAM data while Babyk's sample included both IRAM and ALMA data. Note that only 20 low-mass objects are detected in CO, while the remaining 11 are upper limits. In the case of central cluster galaxies (BCGs) we include 17 upper limits.

\begin{table}
\centering
\caption{ $\rm H_2$ and $\rm HI$ measurements. References: [1] - \cite{Serra:19}, [2] - \cite{Chung:09}, [3] - \citet{Welch:10}, [4] - \citet{McNamara:90}.}\label{tab_sam}
\begin{tabular}{llcc}
\hline
& \\
Name &  $\rm H_2$ & $\rm HI$ & Ref. \\
     & $\times$10$^7 M_{\odot}$ & $\times$10$^7 M_{\odot}$ & \\
& \\
\hline
& \\
NGC596 &3.72 & 14.5 & [3]\\
NGC855 &0.07 & 2.68 &  [3]\\
NGC1052 &3.78 & 4.53 & [3]\\
NGC2768 &4.99 & 19.8 & [3]\\
NGC3073 &0.77 & 16.6 & [3]\\
UGC7354 &0.37 & 10.1 & [3]\\
NGC4283 &0.34 & 4.15 & [3]\\
UGC7767 &0.72 & 8.18 &  [3]\\
& \\
\hline
& \\
NGC1316 & 60.8$\pm$4.9 & 4.4$\pm$0.4 & [1]\\
NGC4374 & 0.57$\pm$0.2 &  $<$1.8 & [2]\\
NGC4261 & $<$4.9 & $<$49.0 & [2]\\
NGC4382 & $<$2.5 & $<$0.9 & [2]\\
NGC4472 & 0.06$\pm$0.01 & $<$0.7 & [2]\\
NGC4552 & $<$1.9 & $<$0.7 & [2]\\
NGC4649 & $<$2.8 & $<$1.5 & [2]\\
& \\
\hline
& \\
     & $\times$10$^{10} M_{\odot}$ & $\times$10$^9 M_{\odot}$ & \\
&\\
\hline
&\\
A1795   &  0.39$\pm$0.04   &$<$5.2 & [4] \\
A1991	&	0.88$\pm$0.05  & $<$3.6    & [4]\\
A2052	&	0.30$\pm$0.03  & $<$3.5    & [4]\\
A2204	&	3.3$\pm$0.12   & $<$1.1    & [4]\\
A2597	&	1.3$\pm$0.4    & $<$4.3    & [4]\\
A262	&	0.39$\pm$0.02  & $<$1.2    & [4]\\
A496	&	0.72$\pm$0.03  & $<$4.7    & [4]\\
Hydra-A	&   1.1$\pm$0.03   & $<$0.84   & [4]\\
A2029	&	1.3$\pm$0.05   & $<$11.0   & [4]\\
AWM7	&	0.28$\pm$0.02  & $<$2.2    & [4]\\
& \\
\hline
\end{tabular}
\end{table}

The situation for HI mass measurements is worse. However, we found 11 HI emission detections, given in  \citet{Serra:19} and \citet{Welch:10} are available. The remaining 16 HI measurements are upper limits \citet{McNamara:90, Chung:09}. We are able to explore the $ \rm H_2/HI-P$ relation for only 17 targets listed in Table~\ref{tab_sam}.  While this number is low, {\bf }{we have assembled enough measurements to explore this relationship in a statistically meaningful fashion}.  

\section{Cold gas mass vs. hydrostatic pressure}\label{sec_res}

\begin{figure}
\centering
\includegraphics[width=0.49\textwidth]{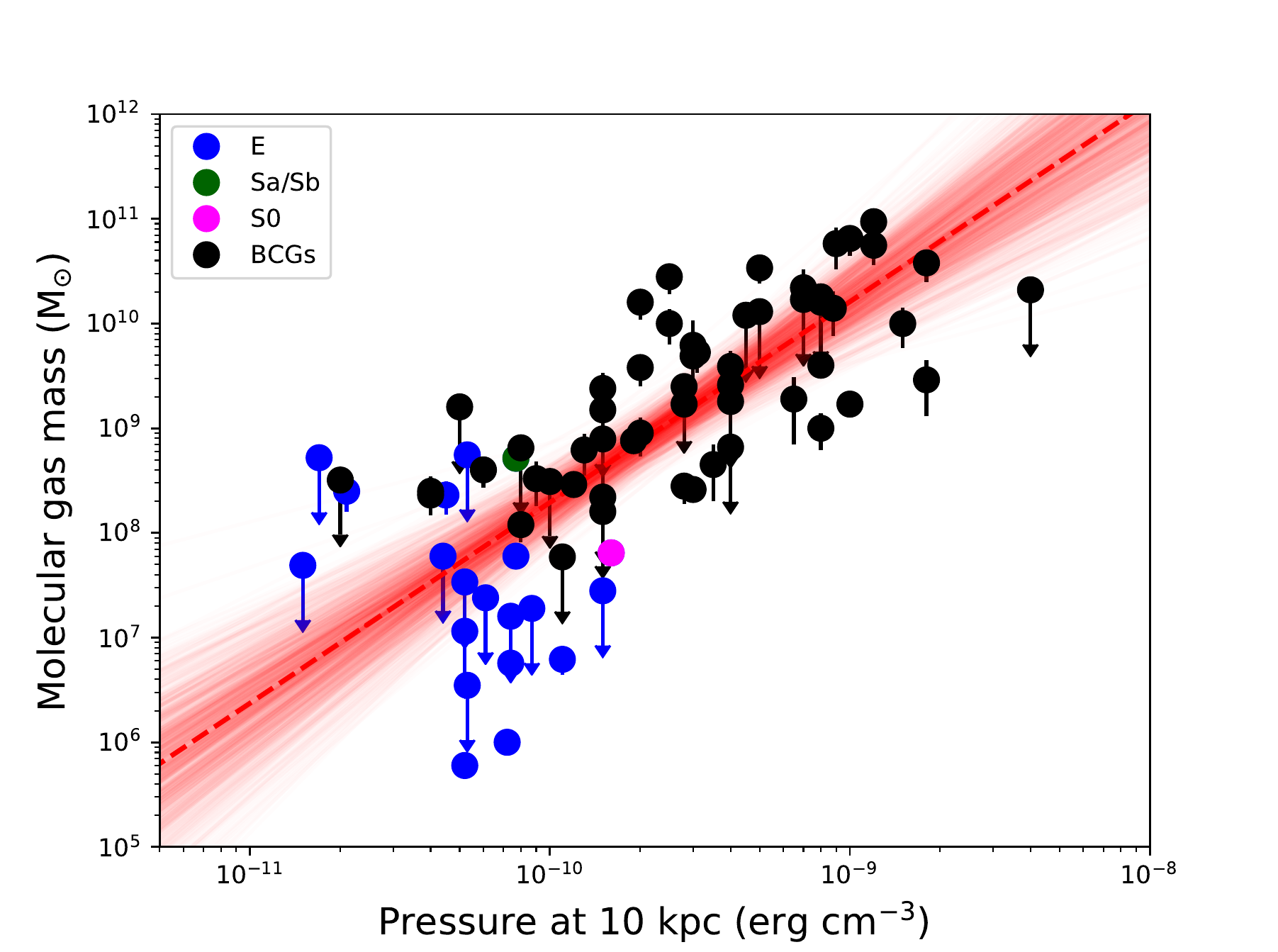}
\caption{The cold molecular gas mass plotted against the pressure at 10 kpc. The red dashed and solid lines represent the best-fit slope and its uncertainties, respectively.}\label{fig1}
\end{figure}

\begin{figure*}
\centering
\includegraphics[width=0.49\textwidth]{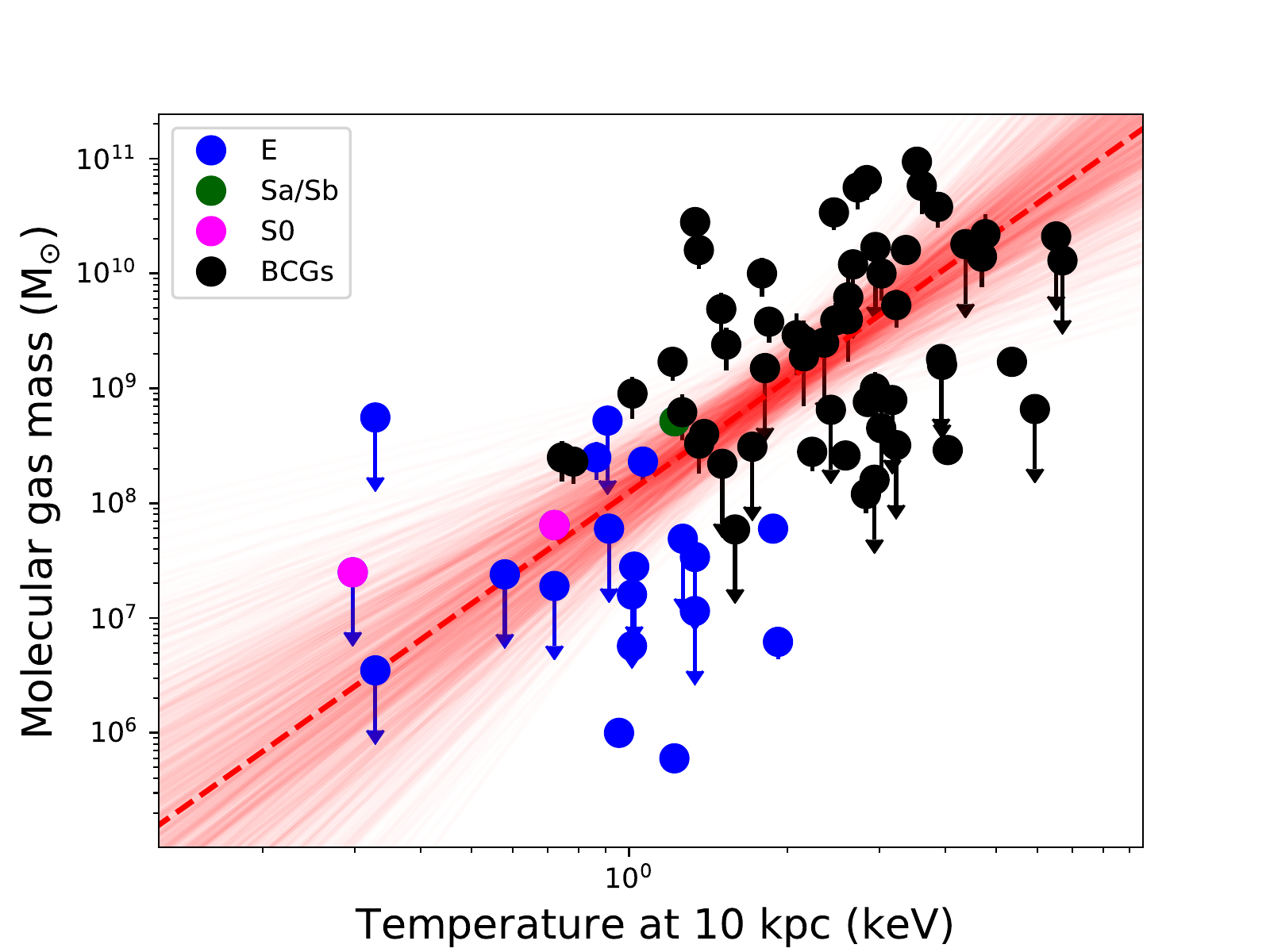}
\includegraphics[width=0.49\textwidth]{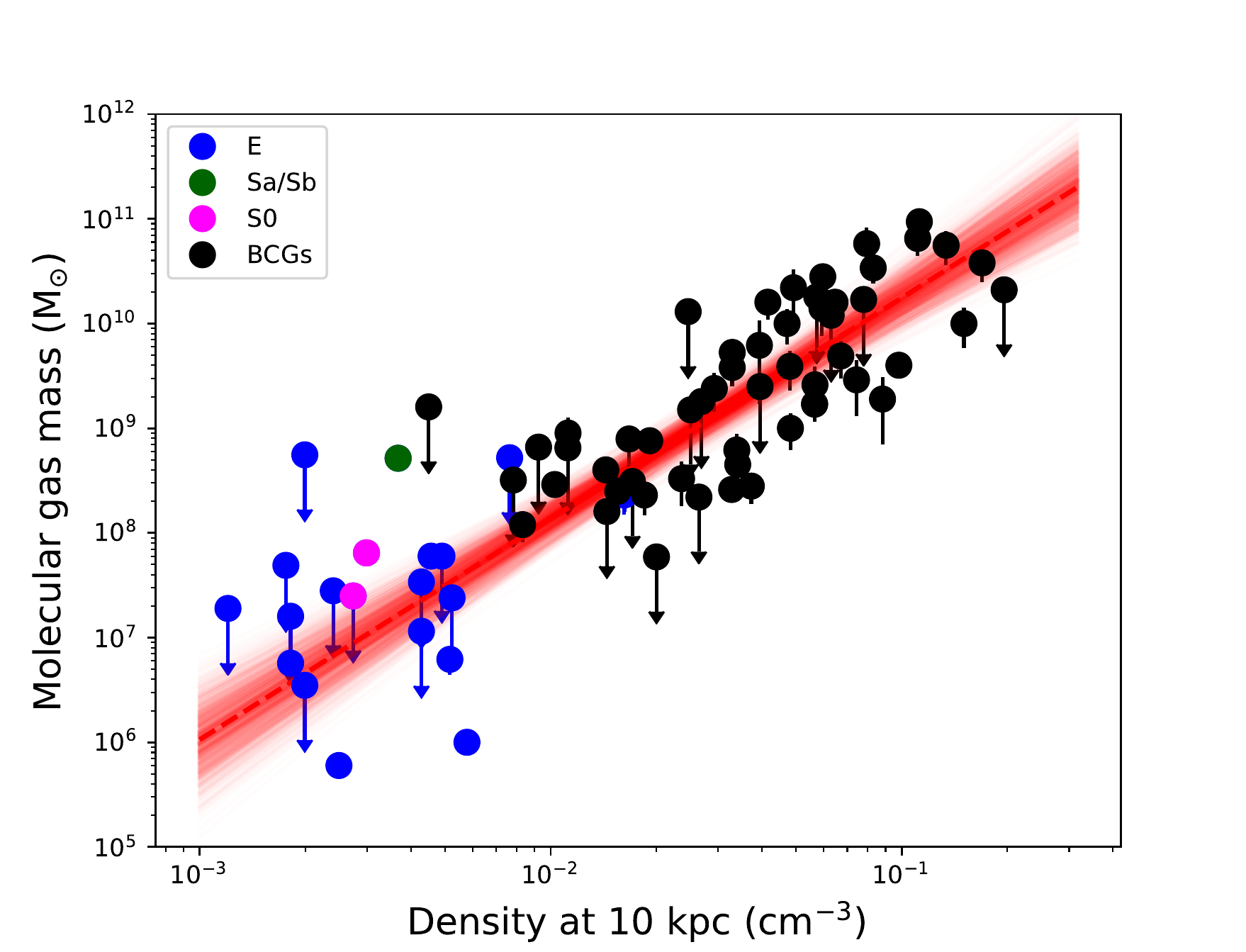}
\caption{The cold molecular gas mass plotted against temperature (left) and electron density (right) at 10 kpc.}\label{fig33}
\end{figure*}

The relationship between molecular gas mass and hydrostatic pressure is shown in Fig.~\ref{fig1}.  Hydrostatic pressure is measured at a characteristic radius of 10 kpc.  Measuring the pressure at 10 kpc ensures the temperature and density measurements are well resolved in all targets over the observed redshift range. {\bf}{ We examined data at radii of 1 and 5 kpc to determine how the trends depend on radius. The inner bin was adopted when resolution prevented us from making a measurement at 1 kpc. All trends remain qualitatively similar and do not depend on the choice of radius, as indicated in Table 2. } The temperature and density correlations shown in Fig.~\ref{fig33}, studied earlier in \citet{Pulido:17} and \citet{Babyk:17prof}, are used to examine the scatter in pressure.  The correlations between pressure and density are tighter than with temperature.  The density trend is the tightest.  The increased scatter in the pressure plot is due to the scatter in temperature, which in turn is related to halo mass.  Applying the likelihood-based approach of \citet{Kelly:07}, the $M_{mol}-P$ relation follows a power-law with a slope of 1.9$\pm$0.3. 
The uncertainties of the best-fit parameters were calculated by running 15000 iterations of MCMC. We use several statistical approaches to test this relation for normality. Fit parameters including slopes and normalizations, uncertainties, $p$-values for null hypotheses, and intrinsic scatter are given in Table~\ref{tab_res}.  The plot is shown in Fig.~\ref{fig1}.  A similar slope of 1.9 is found for density vs molecular gas mass \citep{Babyk:17prof}.  

Both the pressure and density trends with molecular gas mass may be understood in the context of cooling from the hot atmosphere. For an ideal hot system, the cooling rate of the atmosphere scales as $\dot M \propto \rho_{Atm}^2 ~T^{1/2} \propto P^2~T^{-3/2}$, where $\rho_{Atm}$ is the atmospheric electron density and $P$ is the hydrostatic pressure. {\bf}{ In cooler systems, like most of those discussed here, cooling scales quadratically with density and pressure but with temperature scaling as $\dot M \propto \rho_{Atm}^2 ~T^{-1/2} \propto P^2~T^{-5/2}$. Therefore, both density and pressure should scale almost quadratically with the total cold gas mass.  This scaling ignores metallicity variations and assumes unimpeded cooling which is unlikely to hold over the lives of these systems. Nevertheless, the trend in \label{fig_1} is roughly consistent with this scaling with gas density and pressure being the dominant factors.} Molecular clouds presumably accumulated from net cooling over the past $\sim \rm Gyr$ or so.  
 Cooling is almost certainly the origin of the molecular gas in cluster central galaxies.  Its origins in lower-mass atmospheres of early-type galaxies,  which have more scatter about their scaling relations \citep{Babyk:17prof, Lakh18, Werner19}, are less certain.
Nevertheless, the correlation with pressure for the early-type galaxies is consistent with the cluster centrals and is somewhat tighter than the $M_{mol}-M_{X}$ relation \citep{Babyk:17prof}. 

\begin{table*}
\centering
\caption{$M_{mol}$ vs. $P, kT, n_e$ relations of the form log($y$) = $a$ + $b$ log($x$) for features defined at 1, 5, and 10 kpc.}\label{tab_res}
\begin{tabular}{lccccccc}
\hline
Relation & $a$ & $b$ & $p$-Pearson & $p$-Spearman & $p$-AD & $p$-SW & Intr.  scatter \\
\hline
 &&& at 10 kpc &&& \\
 \hline
$M_{mol}-P$  & 13.08$\pm$0.04 & 1.9$\pm$0.3 & $>>$0.0001 & $>>$0.0001 & 0.97 & 1.61 & 0.9$\pm$0.2 \\
$M_{mol}-kT$ & 8.09$\pm$0.27 & 3.24$\pm$0.78 & $>>$0.0001 & $>>$0.0001 & 0.95 & 1.33 & 1.3$\pm$0.3\\
$M_{mol}-n_e$ & 12.89$\pm$0.25 & 2.35$\pm$0.15 & $>>$0.0001 & $>>$0.0001 & 0.98 & 1.22 & 0.4$\pm$0.1\\
$HI-P$ & 4.99$\pm$1.61 & 1.76$\pm$0.89 & $>>$0.0001 & $>>$0.0001 & 0.95 & 1.43 & 0.4$\pm$0.2 \\
$H_2/HI-P$ & 0.83$\pm$0.06 & 1.09$\pm$0.16 & $>>$0.0001 & $>>$0.0001 & 0.98 & 1.24 & 0.3$\pm$0.1\\
\hline
&&& at 1 kpc &&& \\
\hline
$M_{mol}-P$  & 24.12$\pm$1.95 & 1.7$\pm$0.2 & $>>$0.0001 & $>>$0.0001 & 0.99 & 1.75 & 0.7$\pm$0.2 \\
$M_{mol}-kT$ & 8.19$\pm$0.20 & 3.22$\pm$0.50 & $>>$0.0001 & $>>$0.0001 & 0.97 & 1.42 & 0.8$\pm$0.2\\
$M_{mol}-n_e$ & 11.04$\pm$0.38 & 2.15$\pm$0.33 & $>>$0.0001 & $>>$0.0001 & 0.97 & 1.74 & 0.7$\pm$0.2\\
\hline
&&& at 5 kpc &&& \\
\hline
$M_{mol}-P$  & 26.51$\pm$2.50 & 1.9$\pm$0.3 & $>>$0.0001 & $>>$0.0001 & 0.96 & 1.51 & 0.8$\pm$0.2 \\
$M_{mol}-kT$ & 8.08$\pm$0.23 & 3.35$\pm$0.58 & $>>$0.0001 & $>>$0.0001 & 0.98 & 1.53 & 0.9$\pm$0.3\\
$M_{mol}-n_e$ & 11.11$\pm$0.27 & 2.26$\pm$0.20 & $>>$0.0001 & $>>$0.0001 & 0.99 & 1.38 & 0.5$\pm$0.1\\
\hline
\end{tabular}
\end{table*}

\section{ $\rm H_2/HI$ Mass Ratio Dependence on Hydrostatic Pressure}

The link between cold gas mass and atmospheric cooling applies to the total cold gas mass including the atomic and molecular phases.  The prevalence of molecular hydrogen indicates that most of the cold gas is molecular.  This may be related to high pressures at the bases of hot atmospheres, as it is in spirals.  

\begin{figure*}
\centering
\includegraphics[width=0.49\textwidth]{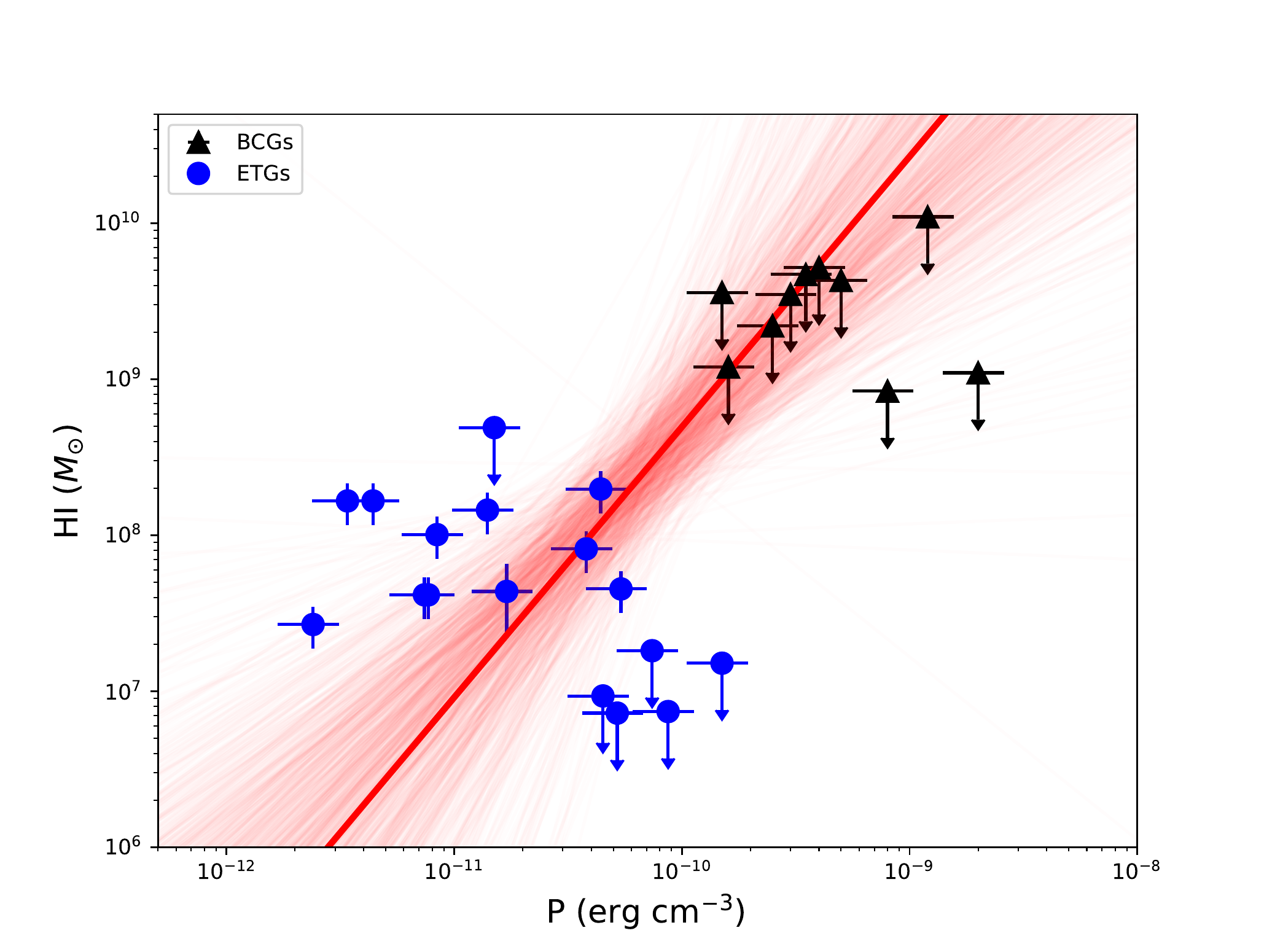}
\includegraphics[width=0.50\textwidth]{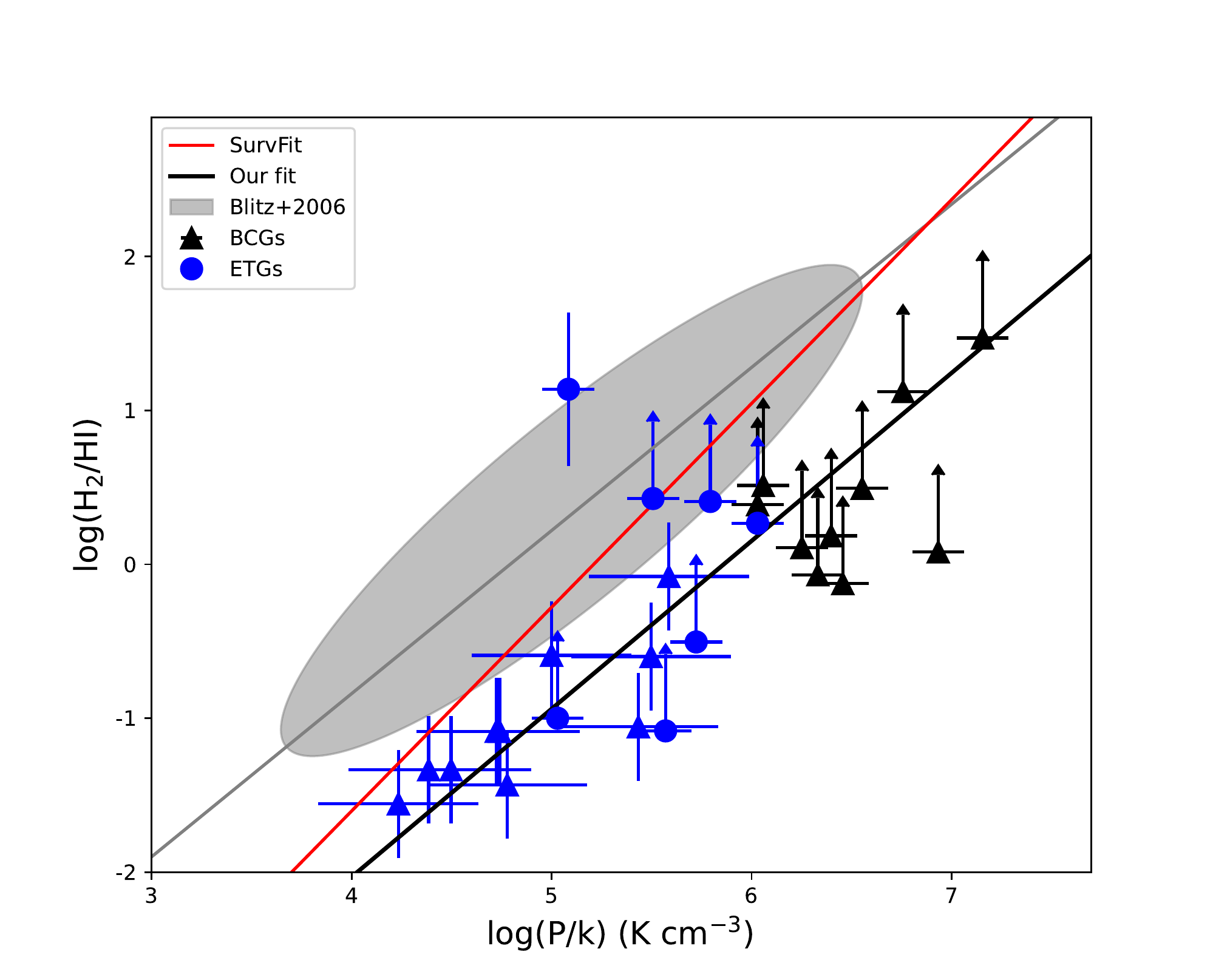}
\caption{Left: The atomic gas mass plotted against hydrostatic pressure at 10 kpc. Right: The ratio of cold molecular gas mass to neutral atomic gas mass plotted against hydrostatic pressure.}\label{fig_2}
\end{figure*}

The relationship between the neutral atomic gas mass and hydrostatic pressure at 10 kpc is shown in Fig.~\ref{fig_2}. The left panel shows HI mass vs pressure in cgs units, while the right panel shows the mass ratio of $H_2/HI$ for systems with both CO and HI observations taken from the literature. In the case of $HI-P$ relation, we got a biased fit with a slope of 1.76$\pm$0.89 and intrinsic scatter of 0.4$\pm$0.2. Although the scatter of data points is large, especially at the low end, the trend between atomic neutral gas and hot gas pressure is obvious. In the case of the $H_2/HI-P$ ratio, the picture is not so obvious due to the high number of lower limits. Thus, we are unable to explore this ratio in detail. However, the lower limits may be meaningfully compared to spiral galaxies with measurements of both gas phases. The range of neutral gas mass limits for both clusters and early-type galaxies is quite large, spanning 10$^7$-10$^{10}~M_\odot$.  This range is comparable to the observed range of molecular gas masses. 

The right panel of Fig.~\ref{fig_2}, shows the $H_2/HI$ mass ratio vs pressure.  Here the pressure is expressed in units of Kelvin cm$^{-3}$ to compare to spiral galaxy midplane pressures from the literature. The gray region and solid line represent the sample and best-fit result of \citet{Blitz:06} and \citet{Leroy:08}, respectively, for spiral galaxies. The ratio $H_2/HI$ should be expressed in surface density units rather than mass.  However, the upper limits and unresolved detection do not permit the surface densities of HI and $H_2$ to be measured. We assume then that the HI and $H_2$ are nearly cospatial, which may not always be true. 

The lower limits for $H_2/HI$ mass are also consistent with the dependence on midplane pressure found for spirals by Blitz and Leroy. Our measurements were fit with a single power-law (R(H2/HI) = Intercept (P/k)$^{Slope}$ by applying (1) Gaussian statistics similar to the $M_{mol}-P$ relation above and (2) by using survival analysis. Survival analysis accounts for uncertainties in both detections and upper/lower limits in order to estimate the form and likelihood of a relationship \citep{statsu, statsu1, statsu2}. In the case of (1) fit we used detections only and got the relation with a slope of 1.09$\pm$0.16 and intrinsic scatter of 0.3$\pm$0.1. In the case (2) fit, we used both detections and lower limits and got 0.61$\pm$0.30 for 3$\sigma$ confidence level. A detailed explanation of the survival statistics analysis is given below.

\begin{figure}
\centering
\includegraphics[width=0.5\textwidth]{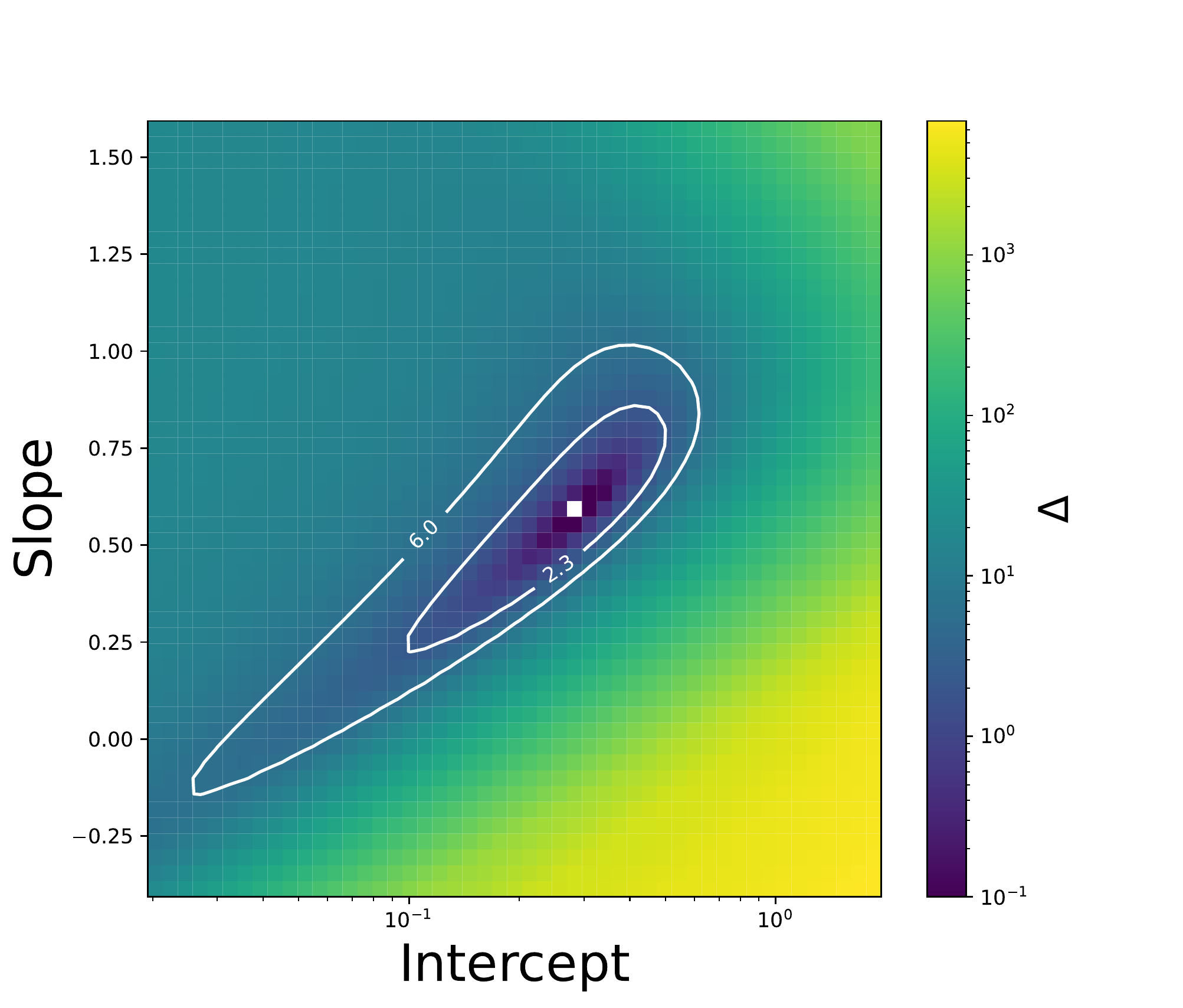}
\caption{Likelihood map for the calculated slope and intercept in Fig. 3 right. $\Delta$ is the likelihood error; the white curves correspond 1 and 3$\sigma$ confidence levels.}\label{fig2d}
\end{figure}

The likelihood is given as
\begin{equation}
 L = \prod_{i=1}^n {\rm Prob}[t_i,\delta_i] = \prod_{i=1}^{n} [f(t_i)^{\delta_i}][1 - S(t_i)]^{1-\delta_i},
 \label{eq1}
\end{equation}
where, $t_i = min(x_i,c_i)$, $x_i$ are the detected values, $c_i$ are the upper limits, $\delta_i$ is 0 for upper limits and 1 for detections. The $f_i$ and $S_i$ are the likelihoods for detections and upper limits, respectively. In the case of Gaussian error distribution, the likelihood of detections is given by
\begin{equation}
 f(x) = \frac{1}{\sqrt{2\pi}\sigma} \exp{\left[ -\frac{1}{2} \left( \frac{x-\mu}{\sigma}\right)^2 \right]},
\end{equation}
where, $x$ and $\sigma$ are the detected values of molecular gas mass and their errors, while $\mu$ is the model. The likelihood for upper limits $S(x)$ is 
\begin{equation}
 S(x) = 1-\frac{1}{2} \left[ 1 + {\rm erf} \left( \frac{x-\mu}{\sqrt{2}\sigma} \right) \right],
 \label{eqsx}
\end{equation}
where $\rm erf$ is the error function, $\rm erf(x) = \frac{2}{\sqrt{\pi}} \int_0^x{e^{-t^2}}dt$. In the case of Eq.~\ref{eqsx} we use $x = ({\rm upper\ limit})_i$ and $\sigma = ({\rm upper\ limit}){_{i={1, 2, 3,.}}}$, where 1, 2, 3, .. indicate upper limits at the $1 \sigma, 2\sigma, 3 \sigma$.. confidence intervals, respectively. 

The results of our two fits are shown in the right panel of Fig.~\ref{fig_2}. The black line represents (1) fit, while the red solid line shows (2) fit. The best-fit (2) parameters are guided by the first term in Eq.~\ref{eq1}. The uncertainties of survival analysis are presented in Fig.~\ref{fig2d}. We tested the slope and intercept by varying parameters while integrating the likelihood equation. We found no significant changes in the slope during the tests. The estimated slope is stable with a value 0.6$\pm$0.3 for 3 sigma confidence. This value is steeper contrary to \citet{Blitz:04, Blitz:06} found the disc midplane pressure scales with the molecular and atomic gas ratio as 
\begin{equation}
    R_{H_2} = \left(\frac{\sum_{H2}}{\sum_{HI}} \right) = \left(\frac{P_m}{P_0}\right)^{\alpha}. 
\end{equation}
Here, ${\sum_{H2}}$ and ${\sum_{HI}}$ are the H$_2$ and HI surface densities, respectively. $P_m$ is the midplane pressure of the disk \citep{Blitz:06, Krumholz:12}. $P_0$ and $\alpha$ are the free power-law parameters. \citet{Blitz:06} found that $\rm H_2/HI$ $\propto$ P$^{0.92}$. $P_0= 2.3\times 10^4~(\rm cm^{-3}~k)$ is the pressure above which the ratio of $H_2/HI$ mass surface densities exceeds unity. All cluster and early-type galaxy systems examined here lie well above this pressure. 

Current instrumentation struggles to detect HI in these systems which hampers our ability to achieve a more detailed picture. Nevertheless, (1) fit shows agreement with the slope obtained by \citet{Blitz:06}, while the (2) fit shows poor agreement with the fit for spirals, but still agrees within errors. The $HI$ upper limits emphasize the key point that these systems are depleted in $HI$ relative to $H_2$. 


\subsection{Discussion}

Central galaxies have yielded high-fidelity measurements of star formation, gas in all phases, and remarkably precise measurements of AGN feedback power.  They are iconic systems that have provided the deepest insights into the complex cycles of feedback and galaxy-black hole co-evolution that likely occur in all massive galaxies. The systems studied here are experiencing radio/mechanical feedback that prevents the atmospheres from cooling catastrophically.  Atmospheric pressure is playing a significant role.  However, much is left to be done.  This includes exploring the role of metallicity in the formation of molecular clouds in these systems.  It further includes understanding how star formation and its accompanying ultraviolet radiation affect the trends and providing scaling relations useful in galaxy formation models.  

Analytic models of molecular cloud formation are governed significantly by gas density and gas metallicity \citep{Krum:08}.   Furthermore, \citet{Ostriker:10} suggested that heating of the interstellar gas plays a critical role in controlling both $H_2$ formation and star formation. In their model, the thermal pressure adjusts to match the midplane pressure set by gravity in disk galaxies.
This coupling may then explain the tight correlation between midplane pressure and higher ratios of $H_2$ to H1 surface density in late-type galaxies \citep{Blitz:04, Blitz:06}.  In our systems, the thermal pressure is also governed largely by gravity and so is broadly consistent with this idea.


The constraints provided here may be useful in modeling galaxy formation \citep[e.g.,][]{Pop:14}.  The role of feedback is crucial and well-studied in our systems and thus may provide insight into how galaxies and their nuclear black holes form and evolve.
Radio-mechanical feedback incorporated into semi-analytic models and simulations have successfully reproduced several properties 
including the galaxy luminosity function \citep{Bower:06, Croton:06}, the average star formation rate and metallicity over time \citep{Monaco:07, Robertson08}, the mix of cold gas in galaxies and its pressure and metallicity dependence \citep{Krum:08,Obre:09,Fu:12,Lagos:11}, and the suppression of cooling atmospheres by AGN  \citep{deLucia:07}.  The work presented here takes the work of Blitz \& Rosolowski to higher masses and pressures found in early-type galaxies without disks. 

Recently, \citet{Morselli:2020} analyzed the HI and H2 content of five nearby, massive main sequence galaxies and linked the availability of molecular and neutral hydrogen to the star formation rate of each region. They found that H$_2$/HI increases with gas surface density, and at fixed total gas surface density it decreases (increases) for regions with a higher (lower) specific star formation rate. \citet{Morselli:2021} studied the evolution of the H$_2$/HI mass ratio with redshift. They found that H$_2$/HI within the optical radius slightly decreases with redshift, contrary to common expectations of galaxies becoming progressively more dominated by molecular hydrogen at high redshifts. We are unable to explore this relation due to the low redshift range of our sample.

\section{Conclusions}\label{sec_summary}
The relationship between molecular gas mass and hydrostatic pressure for 84 central galaxies in clusters, groups, and early-type galaxies is explored. A trend is found between molecular gas mass and hydrostatic pressure measured at an altitude of 10 kpc. The trend follows the power-law slope 1.9$\pm$0.3. This trend with pressure and a similar trend with atmospheric gas density are broadly consistent with the molecular gas having cooled and condensed from the hot atmospheres. We examined the dependence of the $H_2/HI$ mass ratio with pressure for objects with HI measurements. This ratio exceeds unity in all systems, consistent with ratios found in high-pressure regions of spiral galaxies. Our analysis is consistent with a picture where molecular clouds form rapidly from atomic hydrogen in dense, high-pressure atmospheres.  Atmospheric pressure may also serve as a ``valve" that regulates the level of inflow to and outflow from the nuclear black hole \citep{Voit2020}.  Thus pressure may be an important factor determining both the production of nuclear fuel and the level of feedback produced by it.

\acknowledgments
This paper emerged from a conversation between Brian McNamara and Leo Blitz while both were visiting the Flatiron Institute.  Leo was an outstanding scientist and gentleman.  We miss him.

IB acknowledges financial support from the research grant for laboratories/groups of young scientists of the National Academy of Sciences of Ukraine in 2021-22. BRM acknowledges support from the Natural Sciences and Engineering Research Council of Canada, and the hospitality of the Flatiron Institute where this paper germinated. 

This research has made use of data obtained from the Chandra Data Archive and the Chandra Source Catalog, and software provided by the Chandra X-ray Center (CXC) in the application packages CIAO, ChIPS, and Sherpa. We thank all the staff members involved in the Chandra project. Additionally, we have used ADS facilities.

\bibliographystyle{apj}
\bibliography{paper}

\end{document}